\documentclass[aps,prd,twocolumn,showpacs,preprintnumbers,amsmath,amssymb,nofootinbib]{revtex4-1}
\usepackage{amsmath,amsfonts,euscript,amssymb}
\usepackage{epsfig}
\DeclareSymbolFont{bfitletters}{OML}{cmm}{bx}{it}
\DeclareSymbolFont{bfitoperators}   {OT1}{cmr} {m}{n}
\DeclareMathSymbol{\bfitomega}{\mathord}{bfitletters}{"21}

\newcommand{\be}{\begin{equation}}
\newcommand{\ee}{\end{equation}}
\newcommand{\bea}{\begin{eqnarray}}
\newcommand{\eea}{\end{eqnarray}}
\begin{document}

\title{Mixmaster model is associated to Borcherds algebra}

\author{A. E. Pavlov$^{1,2}$}
\affiliation{${}^1$Bogoliubov~Laboratory~for~Theoretical~Physics,~Joint~Institute~of~Nuclear~Research,
~Joliot-Curie~str.~6,~Dubna,~141980,~Russia \\
$^{2}$Institute of Mechanics and Energetics, Russian State Agrarian University, Timiryazevskaya, 49,
Moscow 127550, Russia\\
alexpavlov60@mail.ru}

\begin{abstract}
The problem of integrability of the mixmaster model as a dynamical system with finite degrees of
freedom is investigated. The model belongs to the class of pseudo-Euclidean generalized
Toda chains. It is presented as a quasi-homogeneous system after transformations of phase variables.
An application of the method of getting of Kovalevskaya exponents to the model leads to the generalized
Adler -- van Moerbeke formula on root vectors. A generalized Cartan matrix is constructed with use of
simple root vectors in Minkowski space. The mixmaster model is associated to a Borcherds algebra.
The known hyperbolic Kac -- Moody algebra of Chitre billiard model is obtained by using three spacelike
(without isotropic) root vectors.
\end{abstract}



\maketitle

\section{Introduction}

The problem of initial singularity in cosmology is of particular interest in cosmology during many years
\cite{LifshitzAdv, LifshitzSov}.
Chaotic behavior of various cosmological models in vicinity of the singularity is researched by
different approaches (see, exampli gratia, \cite{IMK, Kirillov, Andrzej, AMMS, MSS, MSchK, KhK}).

The Misner's model \cite{Misner} belongs to the class of pseudo-Euclidean generalized Toda chains \cite{Bog,IM}.
Bogoyavlenskii introduced generalized Toda chains \cite{Bog}. Every simple Lie algebra corresponds to a
completely integrable generalized Euclidean Toda chain. The equations of motion of particles of these chains
admit Lax presentation with $L-A$ pair. There are involutive system of integrals
${\rm tr}\hat{L}^k$, $(k=1,\ldots, n)$.
The ordinary Toda chain corresponds to a simple Lie algebra $\mathfrak{sl}(n, \mathbb{R})$.
A non-periodic chain behaves as asymptotically free \cite{Per}.
The equations of motion of particles in a periodic chain are integrated in theta functions \cite{Adler}.

Adler and van Moerbeke obtained a criterion for the Euclidean generalized Toda chains to be solvable
by quadrature \cite{AvM}. These integrable systems, supplemented the Bogoyavlenskii's solutions, correspond
to Kac -- Moody Lie algebras.
Borcherds extended the class of Kac -- Moody Lie algebras by using a bilinear form which is almost positive
definite \cite{Borcherds}. His algebras generalize Kac -- Moody algebras, adding to the real roots the
imaginary simple roots.

The above examples demonstrate that for understanding the behavior of Toda-like system, it is necessary to
reveal its underlying Lie algebra. In paper \cite{Buyl} there was shown, that in asymptotic billiard
approximation, the algebra is hyperbolic Kac -- Moody one. The model is characterized by chaotic behavior.
The purpose of this work is to find such a generalized Lie algebra that associated to the original
mixmaster Misner's model. This should help us to understand further the behavior of the model,
to obtain a class of functions in which it is naturally described.

\section{Kovalevskaya exponents of mixmaster model}

The equations of motion of some dynamical problems have quasi-homogeneous form.
Let us recall some necessary definitions and theorems for further presentation.
A system of autonomous differential equations
\be\label{systhomo}
\dot{z}_i=v_i(z_1,\ldots, z_n),\qquad 1\le i\le n
\ee
is called quasi-homogeneous \cite{Yoshida} with exponents of homogeneity
$$\lambda_1,\ldots, \lambda_n\neq 0,$$
if there are following identities
\be\label{homogeneous}
v_i (\alpha^{\lambda_1}z_1,\ldots, \alpha^{\lambda_n}z_n)=\alpha^{\lambda_i+1}v_i(z_1,\ldots, z_n)
\ee
for all values $z$ and $\alpha>0$. Thus, the differential equations (\ref{systhomo}) are invariant under
substitutions $z_i\mapsto \alpha^{\lambda_i}z_i, t\to t/\alpha.$
One can differentiate these identities by $\alpha$ and put
$\alpha=1$:
\be
\sum\limits_{j=1}^n \lambda_j z_j\frac{\partial v_i}{\partial z_j}=(\lambda_i+1)v_i.
\ee
Thus the Euler's formula is obtained.

The equations (\ref{systhomo}) have particular solutions:
\be\label{partzi}
z_i=C_it^{-\lambda_i},\qquad 1\le i\le n,
\ee
with coefficients $C_i$, which satisfy the algebraic system of equations
$$
v_i(C_1,\ldots, C_n)=-\lambda_i C_i,\qquad 1\le i\le n.
$$
Equations for variations $\delta z_i$ of the particular solutions (\ref{partzi}) have the following form:
\be\label{vareqns}
\frac{d}{dt}\delta z_i=\sum\limits_{j=1}^n\frac{\partial v_i}{\partial z_j}
(C_1 t^{-\lambda_1},\ldots,C_n t^{-\lambda_n})\delta z_j.
\ee
Differentiating the identity (\ref{homogeneous}) by $z_j$, we obtain
\be\label{alphaij}
\frac{\partial v_i}{\partial z_j}
v_i (\alpha^{\lambda_1}z_1,\ldots, \alpha^{\lambda_n}z_n)=
\alpha^{\lambda_i-\lambda_j+1}\frac{\partial v_i}{\partial z_j}(z_1,\ldots, z_n).
\ee
Substituting $\alpha=1/t$ into the identities (\ref{alphaij}), we can rewrite the system of
differential equations (\ref{vareqns}) as
\be\nonumber
\frac{d}{dt}\delta z_i=\sum\limits_{j=1}^n\frac{\partial v_i}{\partial z_j}
(C_1, C_2,\ldots,C_n) t^{\lambda_j-\lambda_i-1}\delta z_j.
\ee

Taking a solution in the following form
$$\delta z_1=\varphi_1 t^{\rho-\lambda_1},\ldots,\delta z_n=\varphi_n t^{\rho-\lambda_n},$$
and substituting into (\ref{vareqns}), we get a system of linear equations
$$\sum\limits_{j=1}^n\left(K_{ij}-\rho\delta_{ij}\right)\varphi_j=0,\quad i=1,2,\ldots, n.$$
Thus $\rho$ is an eigenvalue, and $(\varphi_1,\ldots,\varphi_n)$ is an eigenvector of the matrix with entries
\be
K_{ij}=\frac{\partial v_i}{\partial z_j}(C_1,\ldots, C_n)+\delta_{ij}\lambda_i.
\ee

The matrix $\hat{K}$ is named Kovalevskaya matrix, and its eigenvalues are Kovalevskaya exponents.
These terms were introduced by Haruo Yoshida, thus marking an outstanding contribution of the Russian woman
to the solution of the important dynamical problem. According to Yoshida's theorem, if a general solution of
the system (\ref{systhomo}) is presented by meromorphic functions in plane of complex time,
then the Kovalevskaya exponents are non-negative integer numbers.
Kovalevskaya has shown that only algebraically completely integrable systems among the
rigid body rotations are Euler's case, Lagrange's case and her famous top \cite{Kov}.
We apply her effective method of investigation to the problem in cosmology.

Let us remind, that $f(z)$ is a quasi-homogeneous function of power $m$ with exponents of quasi-homogeneity
$(\lambda_1, \ldots, \lambda_n)$, if
\be
f(\alpha^{\lambda_1}z_1,\ldots,\alpha^{\lambda_n}z_n)=\alpha^m f(z_1,\ldots, z_n).
\ee
The theorem on the existence of an algebraic integral is proven in paper \cite{Yoshida}.
Let $f$ is quasi-homogeneous integral of power $m$ of the system of equations (\ref{systhomo}) and
$$df(C_1,\ldots, C_n)\neq 0,$$
then $\rho=m$ is the Kovalevskaya exponent.
This important result establishes a connection between the property of meromorphity of a solution and the
existence of integrals. If the system of equations (\ref{systhomo}) has an additional quasi-homogeneous
integral $g$ of the same power $m$, and differentials $df$ and $dg$ are linear independent in point
$(C_1,\ldots, C_n)$, then $\rho=m$ is a Kovalevskaya exponent with multiplicity, more than two.

\section{Mixmaster model as a top of Euler -- Poincar\'e}

An important example for physical applications of homogeneous differential equations are the
Euler -- Poincar\'e equations on Lie algebras \cite{Kozlov}:
\begin{equation}\label{EP}
\frac{dM_i}{dt}=\sum_{j,k=1}^n c_{ik}^j M_j\omega^k,\qquad M_i=\sum_{j=1}^n I_{ij}\omega^j,
\end{equation}
where $1\le i\le n,$
$\bfitomega$ is a vector of angular velocity of the system, ${\bf M}$ is a kinetic momentum,
$c_{ij}^k$ are structural constants of some $n$-dimensional Lie algebra,
$I_{ij}$ is a tensor of inertia of the system considered.
They present a generalization of the known dynamic Euler equations describing
motion of a rigid body with one fixed point. The Lie algebra of a top is the algebra of rotations $so(3).$
The equations (\ref{EP}) possess an integral of energy
\begin{equation}
\label{Energy}
T=\frac{1}{2}\sum_{i,j=1}^n I_{ij}\omega^i\omega^j.
\end{equation}
For homogeneous equations the problem of the uniqueness of the general solution
can be practically brought completely \cite{Yoshida}.

\begin{figure}[tbp]
\begin{center}
\includegraphics[width=2.5in]{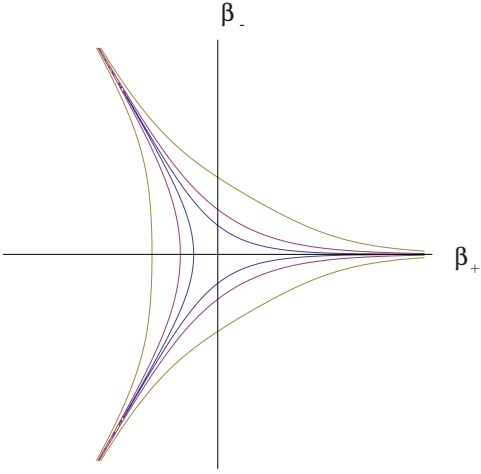}
\caption{\small Lines of level of the potential function $V(\beta_{+},\beta_{-})$. There is a triangular symmetry of the potential and three pockets.}
\label{Contours}
\end{center}
\end{figure}

We show that the mixmaster cosmological model also belongs to the dynamic systems of Euler -- Poincar\'e
on some solvable Lie algebra \cite{Pavlovtop}.
This immediately gives the possibility to use the Yoshida's approach for its analysis.
A superHamiltonian ${\cal H}$ as Hamiltonian constraint of the mixmaster model has the following form:
\be\label{MisnerHam}
{\cal H}=\frac{1}{2}(-p_\alpha^2+p_{+}^2+p_{-}^2)+\exp (4\alpha) V(\beta_{+}, \beta_{-}),
\ee
and a potential function $V(\beta_{+}, \beta_{-})$ is a sum of exponential functions:
\bea
&&V(\beta_{+},\beta_{-})=\exp(-8\beta_{+})+\exp(4\beta_{+}+4\sqrt{3}\beta_{-})+\nonumber\\
&+&\exp(4\beta_{+}-4\sqrt{3}\beta_{-})-2\exp(4\beta_{+})-\label{potential}\\
&-&2\exp(-2\beta_{+}+2\sqrt{3}\beta_{-})-2\exp(-2\beta_{+}-2\sqrt{3}\beta_{-}).\nonumber
\eea
The variables have physical meaning of expansion factor $\alpha$ and anisotropy parameters
$\beta_{+}, \beta_{-}$ of the cosmological model.
The potential $V(\beta)$ is a positive definite potential well with symmetries of an equilateral triangle
in the $\beta_{+} \beta_{-}$ plane (see Fig.\ref{Contours}).
The steepness of the walls of the potential
\bea
V(\beta)&\sim & e^{-8\beta_{+}},\qquad \beta_{+}\to -\infty,\nonumber\\
V(\beta)&\sim & \beta_{-}^2 e^{4\beta_{+}},\qquad \beta_{+}\to +\infty,\qquad |\beta_{-}|<<1\nonumber
\eea
was used to replace the true walls with infinitely hard ones which move in time \cite{Misner}.

Let us pass from canonical Misner's coordinates
$(\alpha,\beta_+,\beta_-; p_\alpha,p_+,p_-)$ to some non-canonical symmetrized variables
$(X,Y,Z;p_x,p_y,p_z)$:
\bea
X&=&\frac{1}{12}\exp (2(\alpha+\beta_++\sqrt{3}\beta_-)),\nonumber\\
Y&=&\frac{1}{12}\exp (2(\alpha+\beta_+-\sqrt{3}\beta_-)),\nonumber\\
Z&=&\frac{1}{12}\exp (2(\alpha-2\beta_+)),\nonumber\\
p_x&=&\frac{1}{12}(2p_\alpha+p_++\sqrt{3}p_-),\nonumber\\
p_y&=&\frac{1}{12}(2p_\alpha+p_+-\sqrt{3}p_-),\nonumber\\
p_z&=&\frac{1}{6}(2p_\alpha-p_+).\nonumber
\eea
Now, the equations of motion are presented as Hamiltonian equations on the direct sum of two-dimensional
solvable Lie algebras
$${\mathfrak g}(6)={\mathfrak g}(2)\oplus {\mathfrak g}(2)\oplus {\mathfrak g}(2):$$
\begin{equation}
\label{algebra}
\{X,p_x\}=X,\quad \{Y,p_y\}=Y,\quad \{Z,p_z\}=Z
\end{equation}
with the superHamiltonian
\bea
\nonumber
{\cal H}&=&-\frac{1}{2}(p_x^2+p_y^2+p_z^2)+\frac{1}{4}{(p_x+p_y+p_z)}^2-\\
&-&2(X^2+Y^2+Z^2)+{(X+Y+Z)}^2. \label{Ham}
\eea
The superHamiltonian (\ref{Ham}) has a form of kinetic energy of a ``top'' (\ref{Energy}):
\begin{equation}
\label{H}
{\cal H}=\frac{1}{2}\sum_{i,j=1}^6 I_{ij}x_i x_j,
\end{equation}
where phase variables are renumbered as follows:
$$x_1=X,\quad x_2=Y,\quad x_3=Z,$$
$$x_4=p_x,\quad x_5=p_y,\quad x_6=p_z,$$
and the tensor of inertia $I_{ij}$ has a block form
\be\label{inertia}
\hat{I}=
\left(
\begin{array}{cccccc}
-2&2&2&0&0&0\\
2&-2&2&0&0&0\\
2&2&-2&0&0&0\\
0&0&0&-1/2&1/2&1/2\\
0&0&0&1/2&-1/2&1/2\\
0&0&0&1/2&1/2&-1/2
\end{array}
\right).
\ee

For the homogeneous system of Euler -- Poincar\'e the Kovalevskaya matrix $\hat{K}$
is built on a partial solution $x_i=C_i/t$ with equal parameters of quasi-homogeneity
$\lambda_i=1$, $(i=1,\ldots, 6)$ (\ref{partzi}).
The components are expressed through the structure constant of the algebra (\ref{algebra})
\begin{equation}
\label{Kov}
K_{ij}=(c_{jk}^i I_{kl}+c_{jk}^l I_{ki})C^l+\delta_{ij},
\end{equation}
and $C_i$ are solutions of the algebraic system
\begin{equation}
\label{Ci}
C_i+c_{ij}^k I_{jl} C_k C_l=0.
\end{equation}
According to calculations implemented initially  in \cite{Pavlovtop},
the spectrum of the matrix (\ref{Kov}) is integer-valued:
$\rho=-1,1,1,2,2,2,$ indicating to the regular behavior of the dynamical system. A negative value $\rho=-1$
is a consequence the system of differential equations to be autonomous.

Particular cases of solutions to the Euler -- Poincar\'e equations on solvable algebras,
branching for any choice of the inertia tensor, were considered in \cite{IK}. Our model does not belong to them.

\section{Generalized pseudo-Euclidean Toda chains}

The Misner's model
belongs to a class of pseudo-Euclidean Toda chains. The superHamiltonian ${\cal H}$ of such class has a form
\be
{\cal H} =\frac{1}{2}<{\bf p},{\bf p}>+\sum\limits_{i=1}^N g_i e^{({\bf a}_i,{\bf q})},
\ee
where a scalar product $<\cdot ,\cdot >$ is introduced in Minkowski space $\mathbb{R}^{1,n-1}$,
$g_i$ are some real coefficients, $(\cdot ,\cdot )$ is a scalar product in Euclidean space $\mathbb{R}^n$;
${\bf a}_i$ we will call as root vectors. Consider a case, when $N\ge n.$
In all known integrable cases momenta ${\bf p}$ and exponential functions are meromorphic functions
of complex time $t$ \cite{Per},
so we will analyze integrability by Birkhoff. Let us set two homomorphisms
$$\mathbb{R}^{1,n-1}\to \mathbb{R}^{1,N-1},\qquad \mathbb{R}^n\to \mathbb{R}^N,$$
introducing $N$ redundant variables, with use of generalized Flaschka mapping
$({\bf p}, {\bf q})\mapsto ({\bf v}, {\bf u})$ \cite{Flaschka}:
$$
v_i=\exp ({\bf a}_i,{\bf q}),\quad u_i=<{\bf a}_i,{\bf p}>,\quad 1\le i \le N.
$$
The equations of motion
\be
\dot{q}_i=\bar\eta_i p_i,\quad \dot{p}_i=-\sum\limits_{j=1}^N g_j a_j^{i}e^{({\bf a}_j,{\bf q})},
\quad 1\le i\le n,
\ee
where the vector $\bar\eta_i (-1,\ldots, 1)$ was utilized, in the new variables $(u_i, v_i)$ are the following
\be\label{systemnew}
\dot{v}_i=u_i v_i,\quad  \dot{u}_i=-\sum_{j=1}^N<{\bf a}_i,{\bf a}_j>g_j v_j,\quad 1\le i\le N.
\ee

The system of equations (\ref{systemnew}) is quasi-homogeneous.
The Poisson bracket of variables $v_i, u_j$
\be
\{u_i, v_j\}=<{\bf a}_i,{\bf a}_j>v_i
\ee
is degenerated in generic case. If there are linear relations between the root vectors
$$
\sum\limits_{j=1}^N\alpha_j{\bf a}_j=0, \qquad \alpha_j\in\mathbb{R},
$$
then
$$F=\sum\limits_{i=1}^N\alpha_i u_i,\qquad \Phi=\prod\limits_{i=1}^N v_i^{\alpha_i}$$
are Casimir functions.

The system of equations has a particular meromorphic solutions:
\be\label{meromorphic}
u_i=\frac{U_i}{t},\qquad v_i=\frac{V_i}{t^2},\qquad 1\le i\le N,
\ee
coefficients of which $U_i, V_i$ subject to a system of algebraic equations
$$V_i(2+U_i)=0,\qquad U_i-\sum\limits_{j=1}^N<{\bf a}_i,{\bf a}_j>g_jV_j=0,$$
where $ i=1,\ldots,N.$

The equations in variations in the neighborhood of the solutions (\ref{meromorphic}):
\bea
\frac{d}{dt}\delta u_i&=&-\sum\limits_{j=1}^N<{\bf a}_i,{\bf a}_j>g_j\delta v_j,\label{deltau}\\
\frac{d}{dt}\delta v_i&=&\frac{U_i}{t}\delta v_i+\frac{V_i}{t^2}\delta u_i\quad 1\le i\le N.\label{deltav}
\eea
Solutions of the system (\ref{deltau}), (\ref{deltav}) we seek in the form:
$$\delta u_i=\xi_i t^{\rho-1},\quad \delta v_i=\eta_i t^{\rho-2},\quad 1\le i\le N.$$

To find the coefficients $\xi_i, \eta_i$ we get a system of linear homogeneous equations with a
spectrum parameter $\rho$:
\bea
(\rho-1)\xi_i&=&-\sum\limits_{j=1}^N<{\bf a}_i,{\bf a}_j>g_j\eta_j,\nonumber\\
(\rho-2-U_i)\eta_i&=&V_i\xi_i,\qquad 1\le i\le N.\nonumber
\eea

Following this way, the formula for the Kovalevskaya exponents $\rho$ has been obtained \cite{PavlovRCh}:
\be\label{AvM}
{\rho=2-2\frac{<{\bf a}_i,{\bf a}_j>}{<{\bf a}_j,{\bf a}_j>}},\quad i\neq j,\quad
{<{\bf a}_j,{\bf a}_j>}\neq 0.
\ee
It generalizes the famous formula of Adler -- van Moerbeke \cite{AvM}, that was obtained for
Euclidean Toda chains. The Cartan matrix of a Kac -- Moody algebra was built
$$\hat{A}\equiv (a_{ij})=\frac{2({\bf a}_i,{\bf a}_j)}{({\bf a}_j,{\bf a}_j)}$$
with use of root systems in Euclidean space \cite{AvM}.
The condition for algebraic integrability of the problem
is the exponents $\rho$ (\ref{AvM}) to be integer. Implementation of this criterion in Euclidean case
leads to linearization of Hamiltonian equations on Abelian varieties \cite{AvM}.
Let us stress, that in our case (\ref{AvM}) the scalar products of root vectors are defined in Minkowski space.

We apply the method to the analysis of integrability of the mixmaster model, root
vectors of which are of the form:
$${\bf a}_1 (4,-8,0),\qquad {\bf a}_2 (4,4,4\sqrt{3}),\qquad {\bf a}_3(4,4,-4\sqrt{3}),$$
$${\bf a}_4 (4,4,0),\qquad {\bf a}_5 (4,-2,2\sqrt{3}),\qquad {\bf a}_6(4,-2,-2\sqrt{3}).$$
The Gram matrix $\hat{G}$ composed of the scalar product of the vectors in Minkowski space is:
\be\label{Gramma}
\hat{G}\equiv <{\bf a}_i,{\bf a}_j>=24
\left(
\begin{array}{cccccc}
2&-2&-2&-2&0&0\\
-2&2&-2&0&0&-2\\
-2&-2&2&0&-2&0\\
-2&0&0&0&-1&-1\\
0&0&-2&-1&0&-1\\
0&-2&0&-1&-1&0
\end{array}
\right).
\ee
Thus we get three root spacelike vectors located outside the light cone $({\bf a}_1, {\bf a}_2, {\bf a}_3)$, and
the other isotropic three ones lie on the light cone $({\bf a}_4, {\bf a}_5, {\bf a}_6)$.
For the first time, it has been shown in paper \cite{IvMel}.
Using the generalized Adler -- van Moerbeke formula (\ref{AvM}),
taking into account zero norm of three vectors, we obtain three equal $\rho=4.$
The classification of semisimple algebras for a case of Euclidean space of the roots was carried out by
Elie Cartan \cite{Per}. However, our matrix (\ref{Gramma}) does not belong to these well studied cases.

\section{Borcherds algebras}

According to Serre's theorem, representations of complex semisimple Lie algebras can be generalized
by constructing a new family of Lie algebras \cite{Kac}.
Cartan matrix associated with some root system can define a new complex Lie algebra.
The structure of an algebra is encoded in its Cartan matrix.
By definition \cite{Kac}, a non-degenerated Cartan matrix is $r\times r$ matrix such that $a_{ii}=2$,
$a_{ij}\le 0$ for $i\ne j$, and $a_{ij}=0$ implies $a_{ji}=0$.
The Cartan matrix $\hat{A}$ sets an algebra of Kac -- Moody $\mathfrak{g} (\hat{A})$.
Generators $h_i, e_i, f_i$ $(i=1,\ldots,r)$ satisfy the Chevalley relations
\begin{eqnarray}
&&[h_i, h_j]=0,\qquad\quad [e_i,f_j]=\delta_{ij}h_j,\label{LieI}\\
&&[h_i,e_j]=a_{ij}e_j,~\quad [h_i,f_j]=-a_{ij}f_j,\label{LieII}
\end{eqnarray}
where $\delta_{ij}$ is the Kronecker symbol.
The matrix $\hat{A}$ is symmetrizable. There exists an invertible matrix $\hat{D}$
with positive elements and a symmetric matrix $\hat{S}$, such that $\hat{A}=\hat{D}\hat{S}$.

A triangular decomposition of ${\mathfrak g}(\hat{A})$ has a form of direct sum of vector spaces
$$
{\mathfrak g}(\hat{A})={\mathfrak n}_{-}\oplus{\mathfrak h}\oplus{\mathfrak n}_{+}.
$$
Here ${\mathfrak h}$ is Cartan subalgebra, which is formed as abelian subalgebra, spanned by the elements $h_i$.
Its dimension $r$ is the rank of the algebra.
The subspaces ${\mathfrak n}_{-}$, ${\mathfrak n}_{+}$
are freely generated. The subalgebra ${\mathfrak n}_{+}$ involves the following commutators
\be\label{multie}
[e_{i_1},[e_{i_2},\cdots ,[e_{i_{k-1}},e_{i_k}]\cdots],
\ee
and the subalgebra ${\mathfrak n}_{-}$ involves commutators
\be\label{multif}
[f_{i_1},[f_{i_2},\cdots ,[f_{i_{k-1}},f_{i_k}]\cdots].
\ee
In generic case, the numbers of multicommutators (\ref{multie}), (\ref{multif}) are infinite.

The Serre relations impose restrictions, which can cut the chains of multiple commutators involving $e_i$
and $f_i$ $(i\neq j)$:
\bea
({\rm ad} e_i)^{1-a_{ij}}e_j&\equiv&[e_i,[e_i,\cdots,[e_i,e_j]]\cdots]=0,\label{Serree}\\
({\rm ad} f_i)^{1-a_{ij}}f_j&\equiv&[f_i,[f_i,\cdots,[f_i,f_j]]\cdots]=0.\label{Serref}
\eea
If $\hat{A}$ is positive, the algebra ${\mathfrak g}(\hat{A})$
is finite-dimensional and falls under the Cartan classification, id est,
it is one of the finite simple algebras $A_n, B_n, C_n, D_n, G_2, F_4, E_6, E_7, E_8.$

If $\hat{A}$ is positive semi-indefinite, id est, $\det \hat{A}=0$ with one zero eigenvalue,
the algebra is infinite-dimensional and is said to be an affine Kac -- Moody algebra.
All affine algebras are classified in
\cite{Kac}. A class of Kac -- Moody algebras, corresponding to the case when the matrix $\hat{A}$
has one negative and other positive eigenvalues is called Lorentzian \cite{Henneaux}.
There is a subclass of the Lorentzian algebras, known as hyperbolic ones.
Kac -- Moody algebra is called hyperbolic if, in addition to Lorentzian,
its Dynkin diagram is that, if removing from it one node one obtains the Dynkin diagram of affine
or finite Kac -- Moody algebra. Lorentzian Kac -- Moody algebras are considered in \cite{Nikulin}.
The classification of hyperbolic root systems is given in \cite{KMclass}.

\begin{figure}[tbp]
\begin{center}
\includegraphics[width=2in]{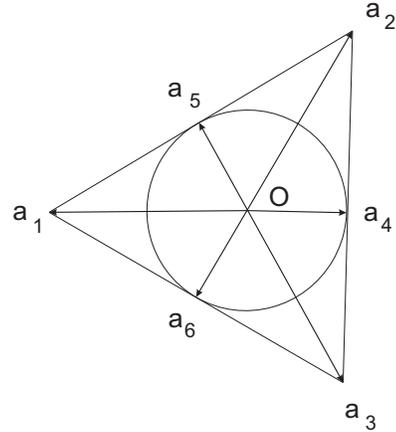}
\caption{\small Top view on a system of the six roots. There is a symmetry of the triangle under rotations on
angles multiple to $2\pi/3$ in the $(\beta_{+}, \beta_{-})$ plane.}
\label{Octonion}
\end{center}
\end{figure}
Borcherds defined generalized Kac -- Moody algebras $\mathfrak{g} (\hat{A})$ by the next generalization of
Cartan matrices \cite{Borcherds}. The symmetrized Cartan matrix is defined as a scalar product of roots
${\bf a}_i$:
$({\bf a}_i, {\bf a}_j)$. Cartan matrix $\hat{A}$ may have non-positive real numbers $a_{ij}\le 0$
on the diagonal and off-diagonal, but all $a_{ij}\in\mathbb{Z},$ if a diagonal element $a_{ii}=2$.
The Borcherds algebra ${\mathfrak g}(\hat{A})$ associated with the generalized Cartan matrix $\hat{A}$
is generated by $3r$ generators ${h_i, e_i, f_i}$ $(i=1,\ldots, r).$
Chevalley commutation relations of Borcherds algebras, corresponding to the generalized Cartan matrix
$\hat{A}$, are analogous to (\ref{LieI}), (\ref{LieII}).
One has to replace the conditions of Serre (\ref{Serree}), (\ref{Serref}) to
$$
({\rm ad} e_i)^{1-2a_{ij}/a_{ii}}e_j=({\rm ad} f_i)^{1-2a_{ij}/a_{ii}}f_j=0,
$$
if a diagonal element $a_{ii}$ is positive. If $a_{ij}=0$, then
$$[e_i,e_j]=[f_i,f_j]=0.$$
Borcherds proved, that his algebras are analogous to the Kac -- Moody ones.
He defined a root ${\bf a}\in\Delta$ to be real, if  $({\bf a},{\bf a})>0;$
otherwise, if $({\bf a},{\bf a})\le 0,$ to be imaginary.

By virtue of the Chevalley relations (\ref{LieII}), the adjoint action of elements $h_i\in{\mathfrak h}$
on $e_i\in{\mathfrak n}_{+}$, $f_i\in{\mathfrak n}_{-}$ is diagonal
\bea
{\rm ad}_{h_i}(e_j)&=&[h_i, e_j]=a_j(h_i)e_j=a_{ij}e_j,\nonumber\\
{\rm ad}_{h_i}(f_j)&=&[h_i, f_j]=a_j(h_i)f_j=-a_{ij}f_j.\nonumber
\eea
Here $a_j(h_i)$ is a linear form on $\mathfrak{h}$.

\section{Kac -- Moody algebra}

The Gram matrix (\ref{Gramma}) is degenerated of rank 3.
Let us notice that the root vectors are divided into three triples
$({\bf a}_4, {\bf a}_2, {\bf a}_3),$ $({\bf a}_5, {\bf a}_1, {\bf a}_2),$ $({\bf a}_6, {\bf a}_1, {\bf a}_3).$
Vectors in an every triple lie in their corresponding plane. The isotropic vectors are obtained as one half of
sums of corresponding spacelike ones:
$$
{\bf a}_4=\frac{1}{2}({\bf a}_2+{\bf a}_3),~~ {\bf a}_5=\frac{1}{2}({\bf a}_1+{\bf a}_2),~~
{\bf a}_6=\frac{1}{2}({\bf a}_1+{\bf a}_3).
$$
Top view on a system of six roots on a Fig.\ref{Octonion} shows a symmetry under their rotations on angles
multiple to $2\pi/3$.
Roots ${\bf a}_4, {\bf a}_5, {\bf a}_6$ lie on a light cone.

Let us restrict our consideration by first three root vectors ${\bf a}_1, {\bf a}_2, {\bf a}_3$
Thus we have a set of simple roots
\be\label{simpleroots}
\Pi({\bf a}_1, {\bf a}_2, {\bf a}_3)\in\Delta_0.
\ee
Spatial components of these vectors are directed along gradients of the dominant wall being done by three terms
of the potential (\ref{potential}). The spatial components of the isotropic vectors are directed to the pockets
of the billiard (see Fig.\ref{Contours}). They do not belong to the lattice spanned on the root system (\ref{simpleroots}).

\begin{figure}[tbp]
\begin{center}
\includegraphics[width=1.5in]{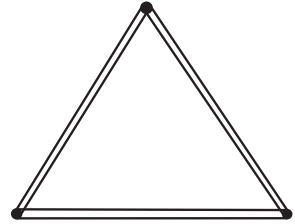}
\caption{\small Dynkin diagram, corresponding to the Kac -- Moody hyperbolic algebra.
There are three nodes and connected them, according to the Cartan matrix $\hat{A}'$, lines.}
\label{Dynkin7}
\end{center}
\end{figure}

The Cartan matrix $\hat{A'}$ is built on the system of simple roots (\ref{simpleroots})
\be\label{Aprime}
\hat{A'}\equiv \frac{2<{\bf a}_i,{\bf a}_j>}{<{\bf a}_j,{\bf a}_j>}=
\left(
\begin{array}{ccc}
2&-2&-2\\
-2&2&-2\\
-2&-2&2
\end{array}
\right).
\ee
The corresponding Dynkin diagram is presented in Fig.\ref{Dynkin7}.
The associated Kac -- Moody algebra is hyperbolic, it has number 7 in the enumeration provided in
\cite{KMclass}.
This result coincides with one that was obtain in the billiard Belinski -- Khalatnikov -- Lifshitz limit
\cite{Buyl},
where the billiard table was identified with the fundamental Weyl chamber of an hyperbolic Kac -- Moody algebra.
Hidden symmetries of the gravitation model were discovered in \cite{DamourH}. The Coxeter group of reflections
is the Weyl group of the infinite-dimensional Kac -- Moody algebra.
It was proven \cite{DHJuliaN,DHNicolai}, if
the billiard region of a gravitational system in BKL--limit can be identified with the fundamental Weyl
chamber of a hyperbolic Kac -- Moody algebra,then the dynamics is chaotic.

\section{Conclusions}

In the present paper it is shown that the cosmological Misner's model as the pseudo-Euclidean generalized Toda
chain beyonds conventional generalized chains. The corresponding algebra is associated to some Borcherds algebra.
As a consequence, it may clear up the question on which class of functions the model is described,
and its behavior should become clear.

Note that all generalized Toda chains associated with simple Lie algebras are integrable due to the presence of
hidden symmetry.
Integrals of motion of a dynamical system are the eigenvalues of a matrix $\hat{L}$,
which depends on the dynamic parameters of the system.
Studies have shown that the correspondence of classical problems with associated Lie algebras is not as obvious
as the connection given by the Emmy Noether's theorem of symmetries.
As different approaches to studying of the model have been demonstrated, the model does not take an extreme
position in general case. According to Yoshida's necessary conditions of integrability, the Kovalevskaya
exponents are non-negative integer numbers.

It should notice, all these methods considered above, are elaborated for obtaining full-parametric
family of solutions. Dynamical systems are presented physical interest, if they are integrated
by Birkhoff conditionally, id est, on zero level of ``energy''.
The method developed can be applied also to study other cosmological Bianchi models.
The factors $g_i$ in front of the exponents of the superHamiltonian parameterize the structure constants
of certain Lie algebra. Changing them, one can get another Bianchi types models.

A gravitational self-dual curvature Riemannian metrics for Euclidean mixmaster model were obtained in \cite{Bel}.
Self-dual Yang -- Mills fields with singularities defined in $S^3$ are obtained in \cite{SelfPavlov}.
All regular solutions none of which is asymptotically Euclidean are classified in \cite{Gib}.
These solutions were used \cite{Latifi} for proofs of nonintegrability of the mixmaster model with
perturbative Painlev\'e test.
If we also consider the Euclidean mixmaster model, substituting the scalar products in (\ref{Aprime})
of the simple roots (\ref{simpleroots})
defined in Minkowski space by their scalar products in Euclidean space, we obtain the following matrix
$$
\hat{A''}\equiv \frac{2({\bf a}_i,{\bf a}_j)}{({\bf a}_j,{\bf a}_j)}=
\left(
\begin{array}{ccc}
2&-2/5&-2/5\\
-2/5&2&-2/5\\
-2/5&-2/5&2
\end{array}
\right).
$$
We see a qualitatively different situation after Wick rotation was implemented:
The matrix is not Cartan's one, because of its off-diagonal entries are fractional.

\section*{Acknowledgments}

I am grateful to participants of the Joint Seminar of the Centre for Gravitation and Fundamental Metrology,
VNIIMS, Institute of Gravitation and Cosmology, Peoples' Friendship University of Russia. I thank to
participants of the seminar of the Bogoliubov Laboratory for Theoretical Physics, Joint Institute of Nuclear
Research for interest to the theme and useful discussion.
Special thanks to Prof. V.D. Ivashchuk for useful numerous discussions.


\end{document}